# Multiwavelength Achromatic Metasurfaces by Dispersive Phase Compensation

Francesco Aieta[1, †], Mikhail A. Kats[1, ‡], Patrice Genevet[1, ⁺] and Federico Capasso[1,*]

[1]School of Engineering and Applied Sciences, Harvard University, Cambridge, MA 02138, USA
†Current affiliation: Hewlett-Packard Laboratories, Palo Alto, CA 94304, USA
‡Current affiliation: ECE Department, University of Wisconsin, Madison, WI 53706, USA
⁺Current affiliation: SIMTech - Singapore Institute of Manufacturing Technology, 638075, Singapore
*capasso@seas.harvard.edu

**The replacement of bulk refractive optical elements with diffractive planar components enables the miniaturization of optical systems. However, diffractive optics suffers from large chromatic aberrations due to the dispersion of the phase accumulated by light during propagation. We show that this limitation can be overcome with an engineered wavelength-dependent phase shift imparted by a metasurface and demonstrate a design that deflects three wavelengths without dispersion. A planar lens without chromatic aberrations at three wavelengths is also presented. Our design is based on low-loss dielectric resonators which introduce a dense spectrum of optical modes to enable dispersive phase compensation. The suppression of chromatic aberrations in metasurface-based planar photonics will find applications in lightweight collimators for displays, and chromatically-corrected imaging systems.**

Refractive and diffractive optical components share many similarities when they are used to manipulate monochromatic light but their response to broadband light is very different. For a material with normal dispersion, refractive lenses have larger focal distances for red light than for blue light and prisms deflect longer wavelengths by a smaller angle; the contrary occurs for diffractive lenses and gratings (*1, 2*). This contrasting behavior arises because two different principles are used to shape the light: refractive optics rely on the phase gradually accumulated through propagation, while diffractive optics operate by means of interference of light transmitted through an amplitude or phase mask. In most transparent materials in the visible the refractive index $n(\lambda)$ decreases with increasing wavelength ("normal dispersion"). Since the deflection angle $\theta$ of a prism increases with $n$ and a lens focal length $f$ is inversely proportional to $n$, the resulting effect is the one shown in Fig. 1A and B. In a diffractive optical element (DOE), the beam deflection angle and the focal length are directly and inversely proportional to $\lambda$, respectively (Fig. 1, C and D), generating an opposite dispersion than the one of standard refractive devices. Although for many applications a spatial separation of different wavelengths is desirable, in many others this represents a problem. For example, the dependence of the focal distance on $\lambda$ produces chromatic aberrations and is responsible for the degradation of the quality of an imaging system. Another difference between these technologies is the efficiency that is generally lower for diffractive optics due to the presence of higher diffraction orders. The wavelength-dependence is typically much more pronounced in diffractive optics than in refractive optics, when low-dispersion materials are used (*2*). In refractive lenses, complete elimination of chromatic aberrations at two and three wavelengths is accomplished using respectively two and three-elements (achromatic doublet and apochromatic triplet) arranged to achieve the same focal length at the wavelengths of interest (*3*). Superachromatic lenses are practically achromatic for all colors by correcting aberrations at four suitable wavelengths (*4*). These strategies, while successful, add weight, complexity and cost to optical systems. Diffractive optical elements on the other hand have the advantage of being relatively flat, light and often low cost. Blazed gratings and Fresnel lenses are diffractive optical devices with an analog phase profile and integrate some benefits of both technologies (e.g. small footprint and high efficiency) but they still suffer from strong chromatic aberrations. Multi-order diffractive lenses overcome this limitation by using thicker phase profiles to achieve chromatic correction for a discrete set of wavelengths (*1*). However the realization of thick, analog phase profiles is challenging for conventional fabrication technologies.

Metasurfaces are thin optical components that rely on a different approach for light control: a dense arrangement of subwavelength resonators is designed to modify the optical response of the interface. The resonant nature of the scatterers introduces an abrupt phase shift in the incident wavefront making it possible to mold the scattered light at will and enabling a new class of planar photonics components (flat optics) (*5-8*).
Different types of resonators (metallic or dielectric antennas, apertures, etc.) have been used to demonstrate various flat optical devices, including blazed gratings (*9-11*), lenses (*12-14*), holographic plates (*15*), polarizers, and wave plates (*6, 16*). The metasurface approach is unique in that it provides continuous control of the phase profile (i.e. from 0 to $2\pi$) with a binary structure (only two levels of thickness), circumventing the fundamental limitation of multiple diffraction orders while maintaining the size, weight, and ease-of-fabrication advantages of planar diffractive optics. However, metasurface-based optical devices demonstrated so far are affected by large chromatic aberrations, though research efforts have shown that relatively "broadband" optical metasurfaces can be achieved (*6, 10-11, 13-15*). This claim of large bandwidth usually refers to the broadband response of the resonators, which is the result of the high radiation losses necessary for high scattering efficiency and, to a lesser extent, of the absorption losses (*6, 17*). As a consequence, the phase function implemented by the metasurface can be relatively constant over a range of wavelengths. However this is not sufficient to eliminate chromatic aberrations. In this report we demonstrate a new approach to planar optics based on metasurfaces that achieves achromatic behavior at multiple wavelengths and offers a potentially practical route to circumventing the limitations of both refractive and standard diffractive optics.

Any desired functionality (focusing, beaming, etc.) requires constructive interference between multiple light paths separating the interface and the desired wavefront (i.e. same total accumulated phase $\varphi_{tot}$ modulo $2\pi$ for all light paths, Fig. 1, E and F). The total accumulated phase is the sum of two contributions: $\varphi_{tot}(r,\lambda) = \varphi_m(r,\lambda) + \varphi_p(r,\lambda)$, where $\varphi_m$ is the phase imparted at point $r$ by the metasurface, $\varphi_p$ is the phase accumulated via propagation through free space and $\lambda$ is the wavelength of light. The first term is related to the scattering of the individual metasurface elements and is characterized by a significant variation across the resonance. The second is given by $\varphi_p(r,\lambda) = \frac{2\pi}{\lambda} l(r)$, where $l(r)$ is the physical distance between the interface at position $r$ and the desired wavefront (Fig. 1, E and F). To ensure achromatic behavior of the device (e.g. deflection angle or focal length independent of wavelength), the condition of constructive interference should be preserved at different wavelengths by keeping $\varphi_{tot}$ constant. The dispersion of $\varphi_m$ has to be designed to compensate for the wavelength-dependence of $\varphi_p$:

$$\varphi_m(r,\lambda) = -\frac{2\pi}{\lambda} l(r) \qquad (1)$$

Where $l(r)$ contains information on the device function (i.e. beam deflector (5, 6), lens, axicon (12), etc.). Equation 1 is the cornerstone for the design of an achromatic metasurface. This approach to flat optics features the advantages of diffractive optics, such as flatness and small footprint, while achieving achromatic operation at selected design wavelengths (Fig. 1). As an example of an achromatic metasurface, we demonstrate a beam deflector based on dielectric resonators: while the typical function of a diffractive grating is the angular separation of different wavelengths, we show beam deflection with wavelength-independent angle of deflection $\theta$ for three discrete telecom wavelengths.

The basic unit of the achromatic multi-wavelength metasurface is a subwavelength size resonator designed to adjust the scattered phase at different wavelengths $\varphi_m(r,\lambda)$ in order to satisfy Eq. 1. In this work, coupled rectangular dielectric resonators (RDR) are used as building blocks (18). Fig. 2A shows the side view of the metasurface: a 240 μm-long collection of silicon (Si) RDRs patterned on a fused silica (SiO$_2$) substrate is designed to deflect normally incident light at an angle $\theta_0$ = -17° for three different wavelengths ($\lambda_1$ = 1300 nm, $\lambda_2$ = 1550 nm, $\lambda_3$ = 1800 nm). The target spatially varying phase functions (Fig. 2A) are defined by:

$$\varphi_m(x,\lambda_i) = -\frac{2\pi}{\lambda_i}\sin\theta_0\, x \quad \text{for } i = 1, 2, 3 \qquad (2)$$

We divide the metasurface into 240 subwavelength unit cells of equal width $s$ and for each section we choose two RDRs of fixed height $t$ and varying widths and separation $w_1$, $w_2$ and $g$ so that the phase response follows Eq. 2. Each unit cell, comprising a slot of width $s$ with two RDRs (Fig. 2A), is different from each other and therefore the metasurface is completely aperiodic unlike other gradient metasurfaces (5, 11, Fig. S7A).

Fig. 2B shows the scattering cross section of an isolated unit cell excited with TE polarization. Given a plane wave travelling along the *z*-axis and incident on the unit cell, at large distance from the interface ($\rho \gg \lambda$) the field distribution is given by two contributions: the light diffracted by the slot and the field scattered by the coupled resonators (*19, 20*):

$$\mathrm{E}(\rho) \approx \frac{e^{jk\rho}}{\rho}[a + b(\theta)] \qquad (3)$$

where *a* is the diffraction amplitude proportional to the amount of incident field that does not interact with the resonators and is in phase with the incident light, $\theta$ is the angle between $\rho$ and the *z* axis and $b(\theta)$ is the complex scattering function. Equation 3 is valid in the limit of slot size *s* significantly smaller than the free-space wavelength λ which is not entirely applicable for our feature size; however this approximation is sufficient to demonstrate the concept. The interference described by Eq. 3 makes it possible to adjust the phase values at several wavelengths simultaneously within a large range. This effect can be visualized using the complex field (phasors) representation in Fig. 2C. While *a* is in phase with the incident field, the phase of *b* -- associated with the scattered light due to the resonances of the dielectric resonators -- spans the range ($\pi/2$, $3\pi/2$) (*11, 21*). The vector sum E can thus cover all four quadrants. Note that the scattering cross section $Q_{scat}$ in Fig. 2B used to visualize the resonance of the structure is related to the forward scattering amplitude $b(0)$ by the optical theorem (*20*).

FDTD simulations were performed to optimize the geometry of each unit cell in order to obtain the desired phase response $\varphi_m(x, \lambda)$ and roughly uniform transmitted amplitude (*18*). We also simulated the entire structure and calculated the far field distribution of light transmitted through the interface at several wavelengths. The fabrication procedure involving chemical vapor deposition of amorphous silicon, electron-beam lithography, and reactive ion etching, is described in the supplementary materials (*18*). Both the simulation and the experimental results show the multi-wavelength achromatic behavior of the metasurface (Fig. 3): while the dispersive nature of conventional flat/diffractive optical components would produce an angular separation of the three wavelengths, the angle of deflection at $\lambda_1$, $\lambda_2$ and $\lambda_3$ is the same: $\theta = -17°$. The diffraction order at the opposite side ($-\theta_0$) is completely suppressed confirming that the structure does not present any periodicity and that the steering effect is the result of the phase gradient introduced by the subwavelength resonators (Fig. S7). Figure 3C shows the simulated and measured deflection angles for normal incidence in the entire spectral range from 1150 to 1950 nm. As expected, the device deflects the incident light at angle $\theta_0$ only for the designed wavelengths. The three colored lines in Fig. 3C are the theoretical dispersion curves obtained from Eq. 2 for metasurfaces designed for fixed wavelengths $\lambda_1$, $\lambda_2$ and $\lambda_3$. The overlap of the experimental and simulated data with these curves indicates that wavelengths other than $\lambda_1$, $\lambda_2$ and $\lambda_3$ tend to follow the dispersion curve of the closest designed wavelengths. Based on the same type of resonators with identical number of degrees of freedom ($w_1$, $w_2$ and $g$), we design four different metasurfaces where the number of corrected wavelengths are one, two, four and five respectively; we find that the deflection angles follow the same trend (Fig. S9). These results suggest a viable path toward the creation of a truly broadband metasurface able to suppress chromatic aberrations for a high number of wavelengths with a reduced number of components compared to refractive optics (*3*).

An important requirement for an achromatic optical device is uniform efficiency within the bandwidth (*1*). We measured the intensity at the angular position $\theta = -17°$ as a function of the wavelength from 1100 nm to 2000 nm (Fig. 3D) and observed intensity variations of less than 13% at $\lambda_1$, $\lambda_2$ and $\lambda_3$ and large suppression ratios with respect to the other wavelengths (50:1). These properties suggest that this device can be used as an optical filter with multiple pass-bands (the full-width at half-maximum for each band is about 30 nm (Fig. S10). We also measured the absolute efficiency of the device (total power at $\theta_0$ divided by the incident power) for the three wavelengths, which is 9.8%, 10.3% and 12.6% for $\lambda_1$, $\lambda_2$ and $\lambda_3$, respectively. From the analysis of the FDTD simulations we can understand the origin of the limited efficiency and how it can be improved (*18*).

The design of a flat lens based on RDRs for the same three wavelengths is also presented. This device is functionally equivalent to the bulk refractive lens known as apochromatic triplet or apochromat (3). The parameters $s$ and $t$ are the same as in the previous demonstration and the values of $w_1$, $w_2$ and $g$ for 600 unit cells are chosen so that the target spatially-variant phase functions are (12):

$$\varphi_m(x, \lambda_i) = -\frac{2\pi}{\lambda_i}\left(\sqrt{x^2 + f^2} - f\right) \quad \text{for } i = 1, 2, 3 \qquad (4)$$

where the focal distance $f = 7.5$ mm. Since we are using two-dimensional RDR, the hyperbolic phase gradient is applied only in one dimension imitating a cylindrical lens. The multi-wavelength properties of the lens are demonstrated with FDTD simulations (Fig. 4). As expected we observe good focusing at $f = 7.5$ mm for $\lambda_1$, $\lambda_2$ and $\lambda_3$ (Fig. 4, C, E and G) and focusing with aberrations at other wavelengths (Fig. 4, B, D, F and H). The diameters of the Airy disks at the focal spots are 50, 66 and 59 μm for $\lambda_1$, $\lambda_2$ and $\lambda_3$ respectively, achieving focusing close to the diffraction limit (40, 47, 55 μm, NA=0.05) (Fig. 4I). For the wavelengths close to $\lambda_1$, $\lambda_2$ and $\lambda_3$, the focal distance follows the dispersion curve associated with the closest corrected wavelength (Fig. 4J). Recently it was pointed out that in order to achieve broadband focusing the phase shift distribution of a metasurface should satisfy a wavelength-dependent function, though a general approach to overcome this inherent dispersive effect was not provided (22).

Note that in general the phase function is defined up to an arbitrary additive constant, therefore Eq. 1 can be generalized as:

$$\varphi_m(r, \lambda) = -\frac{2\pi}{\lambda} l(r) + C(\lambda) \qquad (5)$$

For linear optics applications, $C(\lambda)$ can take on any value and thus can be used as a free parameter in the optimization of the metasurface elements. More generally, $C(\lambda)$ can be an important design variable in the regime of nonlinear optics where the interaction between light of different wavelengths becomes significant.

Metasurfaces have potential as flat, thin and lightweight optical components that can combine several functionalities into a single device, making them good candidates to augment conventional refractive or diffractive optics. The multi-wavelength metasurfaces demonstrated here circumvent one of the most critical limitations of planar optical components: the strong wavelength dependence of their operation (focusing, deflection, etc.). These devices could find application in digital cameras and holographic 3D displays where a red-green-blue (RGB) filter is used to create a color image. Multi-wavelength achromatic metasurfaces could also be implemented in compact and integrated devices for nonlinear processes. Our metasurface design is scalable from the ultraviolet (UV) to the terahertz (THz) and beyond, and can be realized with conventional fabrication approaches. Finally the versatility in the choice of the wavelength-dependent phase allows for functionalities that are very different (even opposite) from the achromatic behavior discussed in this paper. For example an optical device with enhanced dispersion (e.g. a grating able to separate different colors further apart) can be useful for ultra-compact spectrometers.

We gratefully acknowledge partial financial support from AFOSR under grant number FA9550-12-1-0389 (MURI), Draper Lab, Inc. under program SC001-0000000731 and NSF under program ECCS-1347251 (EAGER). We thank B. Kress (Google) for insightful remarks and suggestions; we also thank S. Kalchmair, R. Khorasaninejad and J.P. Laine (Draper Lab) for helpful discussions. The fabrication was performed at the Harvard Center for Nanoscale Systems, which is a member of the National Nanotechnology Infrastructure Network. The thin film characterization of the amorphous silicon was done by Accurion GmbH.

**Supplementary Materials**

Materials and Methods
(1.Fabrication of multi-wavelength metasurfaces, 2.Optical properties of the amorphous silicon, 3.Experimental setup)

Supplementary Text
(1.Rectangular dielectric resonator, 2.Analytical model for rectangular dielectric resonators, 3.Coupled rectangular dielectric resonators, 4.Design of the unit cells, 5.Geometrical aperiodicity of the metasurface, 6.Angle of incidence dependence, 7.Efficiency, 8.From multi-wavelength to broadband, 9.Multiband filter)

Figs. S1 to S10

References (*23-29*)

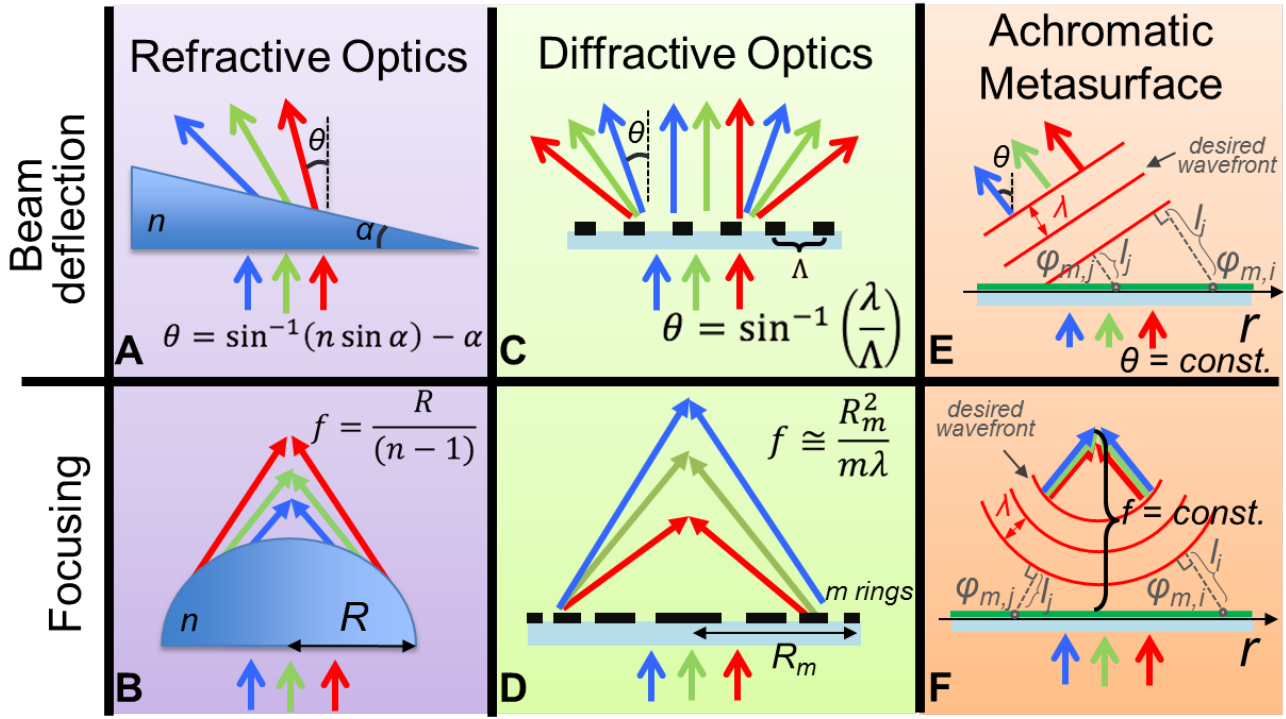

**Fig. 1. Comparison between refractive optics, diffractive optics and achromatic metasurfaces.** In the first two cases (**A-D**) the angle of deflection $\theta$ and the focal length $f$ change as a function of wavelength. The achromatic metasurface (**E**) and (**F**) consisting of subwavelength spaced resonators, is designed to preserve its operation (same $\theta$ and $f$) for multiple wavelengths. In order to achieve this, the phase shifts $\varphi_{m,i}$ and $\varphi_{m,j}$ imparted by the metasurface at points $r_i$ and $r_j$ of the interface, are designed so that the paths $l_i = l(r_i)$ and $l_j = l(r_j)$ are optically-equivalent at different wavelengths.

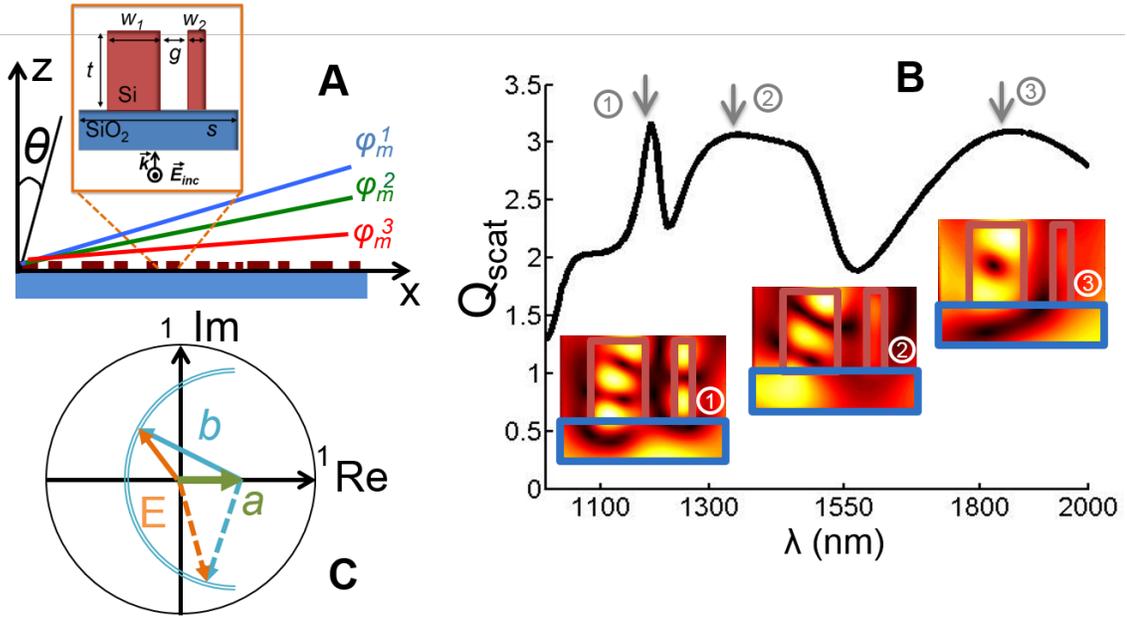

**Fig. 2. Achromatic metasurface.** (A) Side view of the metasurface made of 240 unit cells, each consisting of a slot of the same width $s$, comprising two coupled rectangular dielectric resonators of fixed height $t$ (inset). It is designed to diffract normally incident plane waves at three wavelengths ($\lambda_1$ = 1300 nm, $\lambda_2$ = 1550 nm, $\lambda_3$ = 1800 nm) by the same angle ($\theta_0$ = -17°) by implementing a wavelength-dependent linear phase profile $\varphi_m$. (B) Scattering efficiency $Q_{scat}$ (defined as ratio of the two-dimensional scattering cross-section, which has the dimension of a length, and the geometric length $w_1+w_2$) for one unit cell with geometry $s$=1 μm, $t$=400 nm, $w_1$=300 nm $w_2$=100 nm and $g$=175 nm. The spectrum has resonances due to the individual resonators (2 and 3) and to the coupling between the resonators (1) as shown by the electric field intensity distributions. (C) Vector representation of the interference between the electric fields scattered by the slot and by the resonators, proportional to $a$ and $b$ respectively. The vector sum of $a$ (in green) and $b$ is represented by the phasor $E$ (orange) for two different wavelengths (solid and dashed lines).

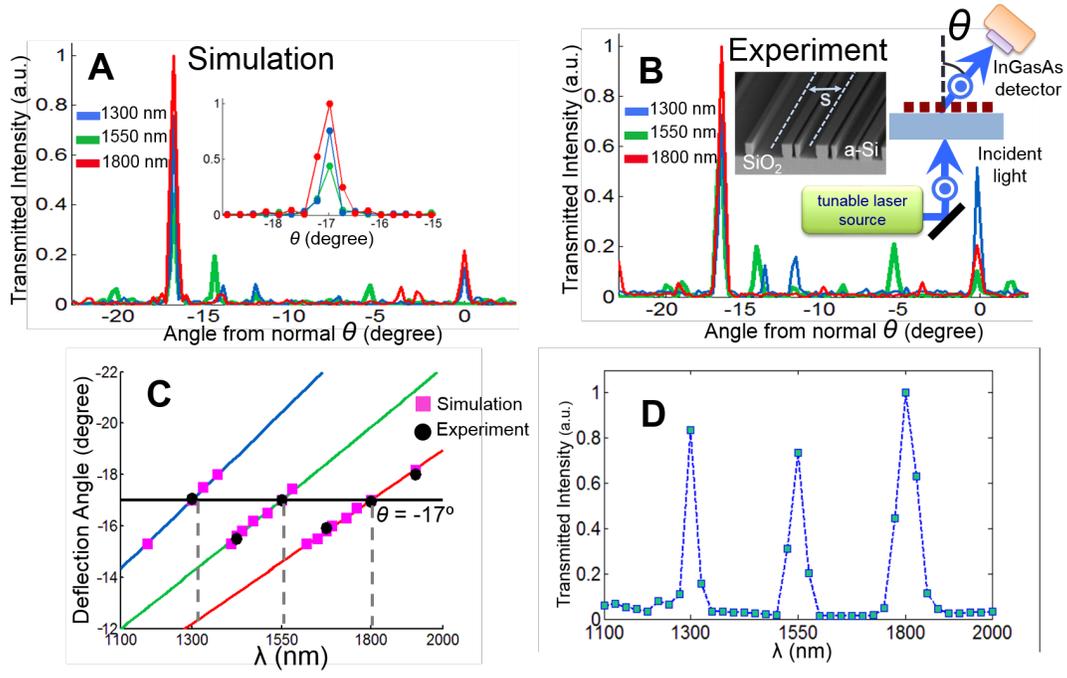

**Fig. 3. Dispersion-free beam deflector.** (**A**) Simulated far field intensity (normalized to the maximum value for each of the three wavelengths) as a function of the angle $\theta$ from the normal to the interface. (Inset) Close-up around the angle $\theta_0$ =-17 deg. (**B**) Far-field transmission measurements. Insets: Schematic of the experimental setup and SEM image of a portion of the metasurface ($s$=1 μm). (**C**) Measured (black circles) and simulated (pink squares) deflection angles for wavelengths from 1100 nm to 1950 nm. The colored lines are calculated from Eq. 2 for fixed phase gradients designed for $\theta_0$=-17° and λ =1300 nm (blue), 1550 nm (green) and 1800 nm (red), respectively. (**D**) Intensity measured by the detector at $\theta_0$ as a function of wavelength.

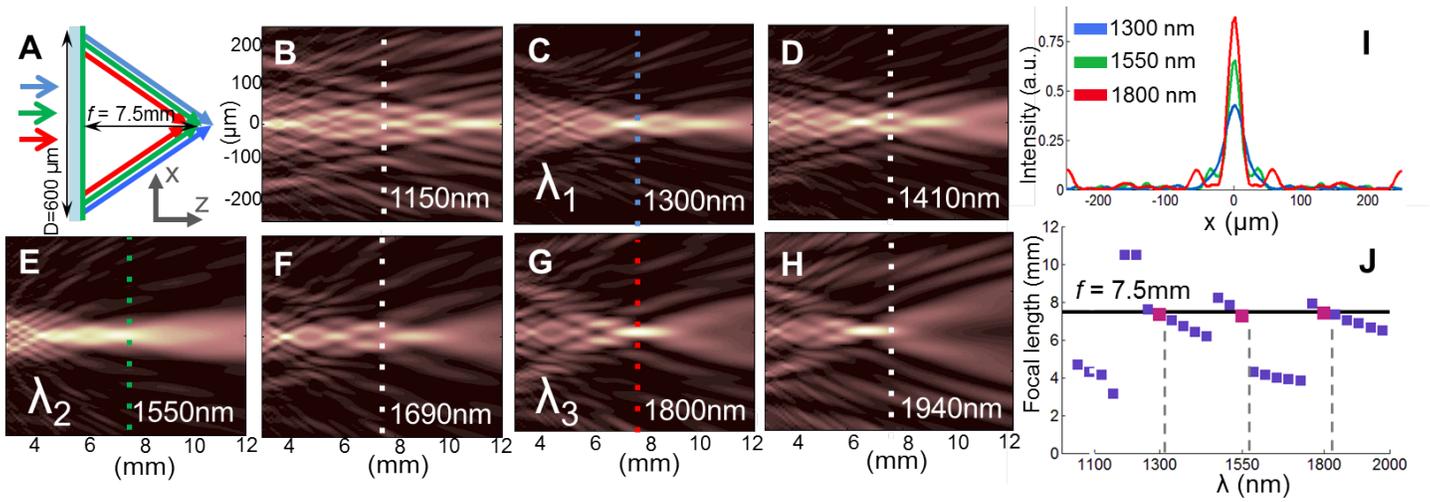

**Fig. 4. Performance of an achromatic flat lens.** (**A**) A broadband plane wave illuminates the backside of the cylindrical lens with side D = 600 μm and focal distance $f$ = 7.5 mm. (**B-H**) Far field intensity distribution for different wavelengths. The dashed lines correspond to the desired focal planes. (**I**) Cross section across the focal plane for $\lambda_1$, $\lambda_2$ and $\lambda_3$. (**J**) Focal lengths as a function of wavelength calculated as the distance between the lens center and highest intensity point on the optical axis. The three pink markers correspond to the wavelengths of interest.